\documentclass[12pt]{iopart}
\usepackage{graphicx}
\usepackage{amssymb}
\begin{document}
\newcommand{\IJMPB}{\textit{Int. J. Mod. Phys. B} }
\newcommand{\PhC}{\textit{Physica C} }
\newcommand{\PhB}{\textit{Physica B} }
\newcommand{\JS}{\textit{J. Supercond.} }
\newcommand{\IEEEmw}{\textit{IEEE Trans. Microwave Theory Tech.} }
\newcommand{\IEEEas}{\textit{IEEE Trans. Appl. Supercond.} }
\newcommand{\IEEEim}{\textit{IEEE Trans. Instr. Meas.} }
\newcommand{\PRB}{\textit{Phys. Rev. B} }
\newcommand{\IJIMW}{\textit{Int. J. Infrared Millim. Waves} }
\newcommand{\APL}{\textit{Appl. Phys. Lett.} }
\newcommand{\JPCS}{\textit{J. Phys. Chem. Solids} }
\newcommand{\AdP}{\textit{Adv. Phys.} }
\newcommand{\Nat}{\textit{Nature} }
\newcommand{\CM}{\textit{cond-mat/} }
\newcommand{\JpnJAP}{\textit{Jpn. J. Appl. Phys.} }
\newcommand{\PhT}{\textit{Phys. Today} }
\newcommand{\ZETF}{\textit{Zh. Eksperim. i. Teor. Fiz.} }
\newcommand{\JETP}{\textit{Soviet Phys.--JETP} }
\newcommand{\EL}{\textit{Europhys. Lett.} }
\newcommand{\Sci}{\textit{Science} }
\newcommand{\EJPB}{\textit{Eur. J. Phys. B} }
\newcommand{\IJMB}{\textit{Int. J. of Mod. Phys. B} }

\newcommand{\eqref}[1]{({\ref{#1}})}

\title[Wideband microwave measurements in Nb/Pd$_{84}$Ni$_{16}$/Nb  and Nb.]{Wideband microwave measurements in Nb/Pd$_{84}$Ni$_{16}$/Nb  structures and comparison with thin Nb films.} 

\author{E Silva$^1$, N Pompeo$^1$, S Sarti$^2$}

\address{$^1$ Dipartimento di Fisica ``E. Amaldi'' and Unit\`a CNISM, Universit\`a Roma Tre, Via della Vasca Navale 84, 00146 Roma, Italy}
\address{$^2$ Dipartimento di Fisica, Universit\`a  ``La Sapienza'', Piazzale Aldo Moro 2, 00185 Roma, Italy}
\ead{silva@fis.uniroma3.it}

\begin{abstract}
We present wideband microwave measurements (1-22 GHz) taken on  Superconductor/Ferromagnet heterostructures with the Corbino disk  technique. We apply the technique to Nb/PdNi/Nb trilayers in the  vortex state and we compare the results to data taken with the same  technique on a Nb thin film. We show that it is possible to directly  extract the genuine flux-flow resistivity from our  frequency-dependent measurements without overcoming the (depinning)  critical current. The characteristic frequency for vortex relaxation  can also be estimated, without resorting to a specific model. We find  that the F layer determines a weakening of pinning and an enhancement  of the flux-flow resistivity with respect to the Nb film. \end{abstract}


\maketitle

\section{Introduction}
\label{intro}

The interplay of superconducting and magnetism, two competing  orderings, has been long studied. A new superconducting state may  arise, the so-called FFLO state, where both the superconducting order  parameters and the magnetic ordering are nonuniform  \cite{larkinJETP65,fuldePR64}. The interplay of superconductivity and  magnetism can be naturally studied in S/F heterostructures (S:  superconductor, F: ferromagnet), where a spatial separation of the  two competing orderings can enhance and somehow control the  interaction between superconductivity and magnetism. The physics of  S/F heterostructures has been deeply studied \cite{buzdinRMP05,  lyuksyutovAdP05} in recent years, and new, interesting aspects were  theoretically predicted and in some cases experimentally confirmed.  As an example, it has been predicted that the proximity-induced  superconducting order parameter does not simply decay in the F layer  as in nonmagnetic metals. Instead, it oscillates in the depth of the  F layer with a scale given by the coherence length in the F layer,  $\xi_F$. This behaviour gives rise to spectacular phenomena, such as  the oscillations of $T_c$ with $d_F$, the thickness of the F layer  \cite{jiangPRL95,kontosPRL02,zdravkovPRL06}, instead of a monotonous  depression of $T_c$ with $d_F$ itself. Also, the vortex state and pinning properties in S/F structures, or  in S films with F particles, dots, pillars, wires etc.  \cite{silhanekAPL06,blamirePRL07, kramerPRL09, aladyshkinSUST09}  received much attention.  

Among the many properties that can give important contributions to  the understanding of the physical background of superconductors, and  thus of S/F heterostructures, stand the microwave response. Indeed,  the temperature dependence of the penetration depth $\lambda$, giving  access to the temperature dependence of the superfluid density  $n_s=\lambda^{-2}(T)/\lambda^{-2}(0)$, has been measured in some S/F  bilayers and S/F/S trilayers: measurements on Nb/CuMn bilayers  \cite{mercaldoPRB99} revealed a temperature dependent superfluid  fraction largely different from the case of superconducting/normal  metal bilayers, suggesting a temperature-independent proximity  induced correlation length. Mn impurities in Nb/Cu bylayers did not  affect the estimated zero-temperature $\lambda$ \cite{palombaPhB00}.  In Nb/Ni bilayers \cite{lembergerJAP08} the temperature dependence of  the superfluid density and the $d_F$ dependence of $\lambda(0)$ were  found to be consistent with significant superfluid density inside the  F layer. While there is certainly need for additional measurements of  the penetration depth, it is surprising that there seem to be no  measurements at all of the vortex state complex resistivity at radio  and microwave frequencies, as opposed to conventional superconductors  \cite{GR,JanjusevicPRB06,songPRB09}, diborides \cite{dulcicPRB03,  sartiPRB05}, cuprates \cite{golosovskySUST96,tsuchiyaPRB01},  nanostructured high-$T_c$ superconductors  \cite{PompeoAPL07,PompeoJAP08}. Such measurements can be particularly  useful since they give direct access to the pinning constant and the  free-flux-flow resistivity (see below). Thus, we started an  experimental investigation of the microwave properties of S/F/S  heterostructures in the vortex state. 

Microwave resonators are a well-established technique to measure with  high resolution the complex resistivity of thin superconducting films  in the vortex state \cite{peligradPRB98, pompeoJSUP07}. However, the  resonators lack the capability to perform measurements in an extended  (and continuous) frequency range. Since no measurements exist of the  vortex state microwave resistivity in S/F heterostructures, it would  be hazardous to rely on single-frequency measurements to draw  conclusions on the usually complicated response in the vortex state.  Thus, we applied to S/F/S trilayers the nonresonant, wideband Corbino  disk technique \cite{boothRSI94,sartiCM04}: the technique is able to  reliably extract the resistivity in a rather wide frequency range,  conservatively between 1 and 20 GHz. It is then possible to compare  the measurements to a wide class of models to get the important  vortex state parameters. 

Before describing with some detail the technique, we briefly recall  the expectations of the frequency-dependent complex resistivity in  the vortex state, $\rho_{vm}$, as calculated within mean field models  with a static magnetic induction field $\mathbf{B}$ perpendicular to  the microwave currents. At microwave frequencies the vortex  displacement due to the force exerted by the alternate currents  $\mathbf{J}_{\mu w}$ is very small \cite{tomaschPRB88}, and it is  customary to write down the equation of forces (per unit length)  acting on a single vortex line in the local limit as follows  \cite{GR, golosovskySUST96,BS, CC}:

\begin{equation}
\label{eq:forces}
     \eta\mathbf{v}+\nabla U=
     \mathbf{J}_{\mu w}\times\mathbf{\hat{n}}\Phi_0+\mathbf{F_{thermal}}
\end{equation}
\noindent where $\mathbf{v}$ is the vortex velocity,  $\mathbf{\hat{n}}$ is the unit vector along the vortex, $\Phi_0$ is  the flux quantum, $\mathbf{J}_{\mu w}\times\mathbf{\hat{n}}\Phi_0$ is  the Lorentz force exerted on the flux line and the stochastic force  $\mathbf{F_{thermal}}$ takes into account thermal effects,  responsible for fluxon jumps between pinning sites. The viscous drag  coefficient $\eta$, which is also commonly referred to as vortex  viscosity, takes into account the power dissipated by moving vortices  due to, e.g., the relaxation processes of quasiparticles and  scattering processes due to currents flowing in the vortex cores  \cite{BS,tinkhamPRL64,kopninRPP, kopninBOOK}. As such, it bears  informations related to the microscopic electronic state. The effects  of pinning are represented by the force $-\nabla U$ where $U$ is the  spatial function describing the pinning potential. The appearance of  a Hall term in \eqref{eq:forces} can be neglected in superconductors  in the dirty limit.

From the basic equation of motion (\ref{eq:forces}), many models can be derived, differing essentially for the impact of thermal effects and elasticity of the vortex lattice \cite{GR, CC, brandt, MSexp}. Nevertheless, it has been shown \cite{pompeoPRB08} that the seemingly large variety of models yield expressions for the vortex resistivity $\rho_{vm}$ that can be cast in a single, very general expression:
\begin{equation}
\label{eq:rhovm}
\rho_{vm}=\rho_{vm,1}+\rmi\rho_{vm,2}=\rho_{ff}\frac{\epsilon_{eff}+\rmi\nu/\nu_{0}}{1+\rmi\nu/\nu_{0}}
\end{equation}
\noindent where the dimensionless parameter $1\geq \epsilon_{eff}  \geq 0$ is a measure of the weight of creep phenomenon, $\nu$ is the  measuring frequency and $\nu_{0}$ is the main characteristic  frequency governing the vortex relaxation phenomena. With respect to  the well-known depinning frequency \cite{GR} $\nu_p$, one has  $\nu_0\simeq \nu_p$ when $\epsilon_{eff}\ll 1$, that is when thermal  effects can be neglected (one should mention that this is the  approximation under which most of microwave measurements in  conventional superconductors have been interpreted in the past). The  flux-flow resistivity $\rho_{ff}$ represents the vortex resistivity  in absence of any pinning and creep effect, that is the  \textit{ideal} vortex flow resistivity. As it can be seen, it is  possible to obtain the true flux flow resistivity by simply  increasing sufficiently the measuring frequency $\nu$, without the  need for (a) ultrapure samples and (b) high transport currents as in  dc, with the related technical problems due to heating.

\begin{figure}[h]
\centerline{\includegraphics[width=10cm]{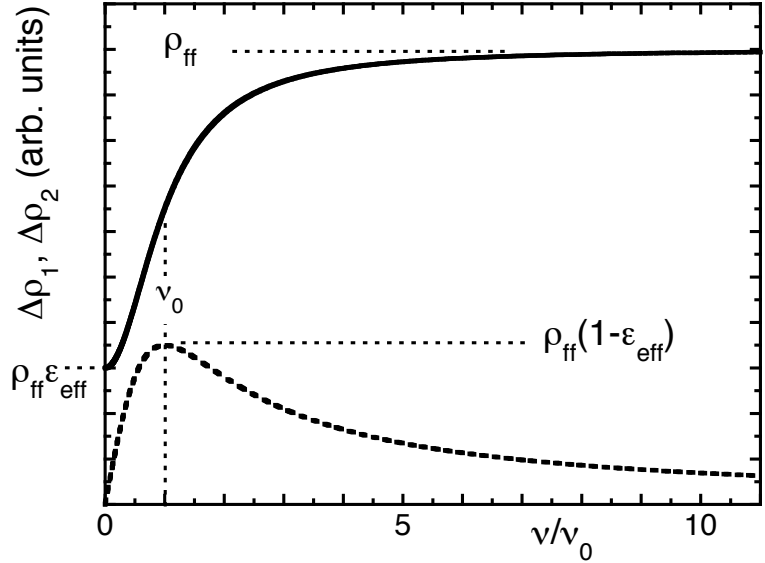}}
   \caption{Complex vortex resistivity $\rho_{vm,1}+\rmi\rho_{vm,2}$  as a function of the reduced frequency $\nu/\nu_0$ plotted on the  basis of Equation \ref{eq:rhovm}. Full line: $\rho_{vm,1}$, dashed  line: $\rho_{vm,2}$. The direct determination of the vortex  parameters $\rho_{ff},\nu_0,\epsilon_{ff}$ is indicated in the  figure.}
   \label{figteo}
\end{figure}

The frequency dependence of $\rho_{vm}$ as described by  (\ref{eq:rhovm}) is depicted in Figure \ref{figteo} as a function of  the reduced frequencies $\nu/\nu_0$ in the range $0.1\leq \nu/\nu_0  \leq12$. As it can be seen, the real part of the vortex resistivity  $\rho_{vm,1}$ contains basically the entire information on the vortex  parameters, $\rho_{ff},\nu_0,\epsilon_{ff}$. Thus, measurements of  $\rho_{vm,1}$ over a sufficiently wide frequency range are of primary  importance.

Aim of this paper is to show how the swept-frequency Corbino disk  technique can be a valuable tool in the study of the vortex state  resistivity, in particular in S/F heterostructures: the little role  played by the models in the analysis of the data, together with the  power of swept-frequency data make the Corbino disk an ideal probe of  new materials where one cannot rely on known properties. We also  present the first measurements of the frequency-dependent vortex  resistivity in two S/F/S trilayers, and we show that differences  emerge with respect to simple Nb even at extremely small $d_F$.

The paper is organized as follows. In the next Section we introduce  the Corbino disk technique, with the expressions linking the measured  quantities and the complex resistivity of superconducting films in  the vortex state. In Section \ref{exp} we briefly describe our  samples, and we present the experimental data for the complex  resistivity taken in a Nb film and two Nb/PdNi/Nb trilayers, with  different thickness of the F layer. We summarize the results and  suggest further developments in Section \ref{conc}.
\section{The Corbino disk technique}
\label{tech}
We have developed in the past the Corbino disk technique  \cite{sartiCM04, tosorattiIJMPB00}, following the pioneering work (in  the context of superconductivity) of the Maryland group  \cite{boothRSI94, WuPRL95}. The technique is conceptually elementary,  consisting in the termination of a coaxial cable with the sample  under study and the measurement of the complex reflection coefficient  $\Gamma$.
In our particular case, we measured the complex  reflection coefficient by means of a Vector Network Analyser, or VNA  (we performed most part of the measurements with an Agilent-HP 8510C,  and a smaller part of the measurements with an Anritsu 37269 D),  connected to a microwave line made up of commercial coaxial cables  with physical cutoff frequency at 60 GHz.
In principle, from the  measured $\Gamma$ one can extract the real and imaginary part of the  complex resistivity of a sample terminating the cable. The practical  realization turns out to be far from being trivial. The technique is  inherently nonresonant, and the reflection coefficient is heavily  affected by losses in the cable, phase stability of the cable itself  and standing wave in the necessarily long microwave line connecting  the sample in the cryostat to the external VNA. 
The achievable sensitivity of the measurement depends  crucially on the calibration and on the characteristics of the VNA  employed. While very small signals can be in principle detected with  a full calibration (by using three known standard loads, typically an  open, a short and a load of known impedance), this is not feasible in  the present case: since both temperature at sample position and  temperature gradients along the cable affect the transmission line,  the conventional calibration is almost impossible to achieve. Thus,  an approximate calibration, with the sample in-situ, has to be used,  as detailed below. We remark that the approximate calibration is the  limiting factor in assessing the sensitivity of the present  measurements. As a rule of thumb, and with some conservative overestimates, in our present setup and with the chosen calibration  procedure no signal below 1\% of the normal state of the  superconducting sample can be detected with sufficient reliability.
 However, the invaluable advantage of swept-frequency measurements  compensates for the reduced sensitivity with respect to, e.g.,  resonator-based microwave measurements.

We described elsewhere \cite{sartiCM04} both the measuring cell and  the issues related to the measurement procedure, with particular  attention to calibration issues. We here recall the basic aspects of  the technique and the specific calibration method here employed.

The reflection coefficient at the sample surface $\Gamma_0$ is related to the impedance of the sample $Z$ through the relation \cite{Collin}

\begin{equation}
\label{gamma-z}
\Gamma_0=\frac{Z-Z_0}{Z+Z_0}
\end{equation}

\noindent where $Z_0$ is the characteristic (wave) impedance of the  dielectric which fills the cable. \footnote{In principle one should  have $Z_0 = 377\;\Omega$ (vacuum impedance) if the last section of  the cable is not filled with the dielectric. In practice, the device  (launcher) actually placed at the terminal section of the cable to  make the contact between the coaxial cable and the sample, does not  include any dielectric but introduces an additional contribution  which in general requires to experimentally determine $Z_0$.} For  thin films (that is, for films whose thickness is lower than the  penetration depth) the sample complex resistivity  $\rho(\nu)=\rho_1(\nu)+\rmi\rho_2(\nu)$ is related to the sample  impedance through the relation $Z=\rho/d$, so that \cite{boothRSI94,  tosorattiIJMPB00}
\begin{equation}
\label{Zs}
\rho(\nu) =  Z_0 d\frac{1+\Gamma_0(\nu)}{1-\Gamma_0(\nu)}
\end{equation}

\begin{figure}
\centerline{\includegraphics[width=10cm]{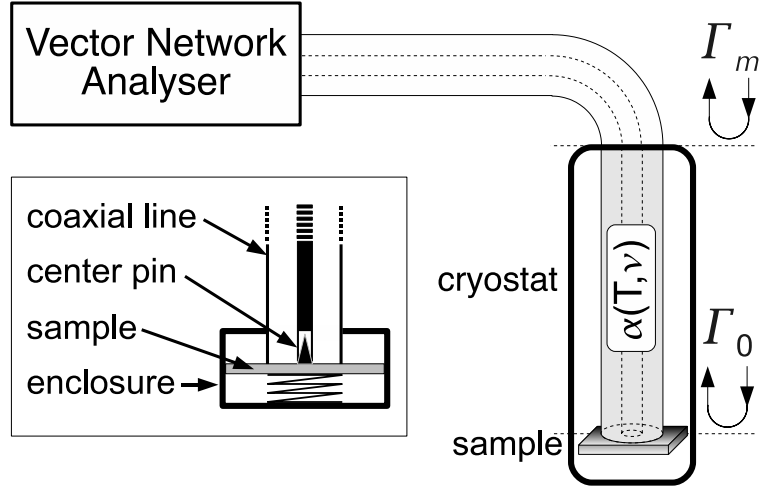}}
\caption{Sketch of the measurement line. Inset: detail of the contact between sample and coaxial line.}
\label{fig:CorbinoLine}
\end{figure}

Unfortunately, a direct measure of $\Gamma_0$ is impossible. One  measures instead (see Figure \ref{fig:CorbinoLine}) the reflection  coefficient at the beginning of the microwave line $\Gamma_m$, which  is related to $\Gamma_0$ through cable coefficients that should be  determined through some calibration procedure.

Since those coefficients depend on temperature and temperature  gradients of the cable, a full calibration is in general not  possible. This crucial problem can be partially overcome  \cite{sartiCM04} by making use of a calibration procedure which  allows to obtain \textit{variations} of $\rho$ with some external  parameter (e.g., the temperature $T$, the applied magnetic field  $H$). For instance, by measuring $\Gamma_m$ at two different  temperatures, it is possible to obtain the sample impedance  difference between the two measured temperatures. This procedure is  based on the hypothesis that most of the cable parameters vary only  smoothly with temperature. This hypothesis can be verified through  the measured data \cite{sartiCM04}. Assuming this hypothesis is  valid, it can be shown that the measured reflection coefficient  $\Gamma_m$ can be related to the reflection coefficient $\Gamma_0$ at  the film surface through the relation

\begin{equation}
\label{alpha}
\Gamma_m(T,H,\nu)\simeq\alpha(T,\nu)\Gamma_0(T,H,\nu)
\end{equation}

The complex coefficient $\alpha(T, \nu)$ describes in a compact way  the residual cable contribution to the measurement. In the present  measurements of the vortex state response, it is essential that this  quantity is field independent: non-magnetic cables and connectors  have to be used.

By focusing on field sweep measurements (i.e. measures performed by  varying the field at fixed $T<T_c$), it is possible to derive  $\alpha(T,\nu)$ by considering data taken at the two distinct fields  $H=0$ and $H^*>H_{c2}$.

At $H^*$, one has ${\Gamma}_m(T, H^*,\nu)=\alpha(T,  \nu)\Gamma_{0,n}=\alpha(T, \nu)(Z_n-Z_0)/(Z_n+Z_0)$, where  $\Gamma_{0,n}$ and $Z_n$ are the  reflection coefficient and surface  impedance of the sample in the normal state. For superconducting  samples in the thin film limit, $Z_n=\rho_n/d$ is a frequency  independent real quantity (since the plasma frequency of the material  is usually much higher than the measuring frequency) lower than  $Z_0$, which is also real. As a consequence, $\Gamma_{0,n}$ is  frequency independent, real and negative.

At $H=0$, one has  $|{\Gamma}_m(T, 0,  \nu)|=|\alpha(T,\nu)\Gamma_0(T,0,\nu)|=|\alpha(T,  \nu)||Z(T,0,\nu)-Z_0|/|Z(T,0,\nu)+Z_0|$. At low frequencies and not  too close to $T_c$, $\Re[Z(T,0,\nu)]\ll Z_0$ so that $|\Gamma_0(T,0,\nu)|\simeq1$  \footnote{Note that this requirement is softer than the similar one,  $\Gamma_0(T,0,\nu)\simeq-1$: the latter involves the additional  requirement $\Im[Z(T,0,\nu)]\ll Z_0$. This is quite useful for the  case of thin superconducting films, where usually $\Re(Z)\ll\Im(Z)$.}.

\begin{figure}[hbt!]
\centerline{\includegraphics[width=10cm]{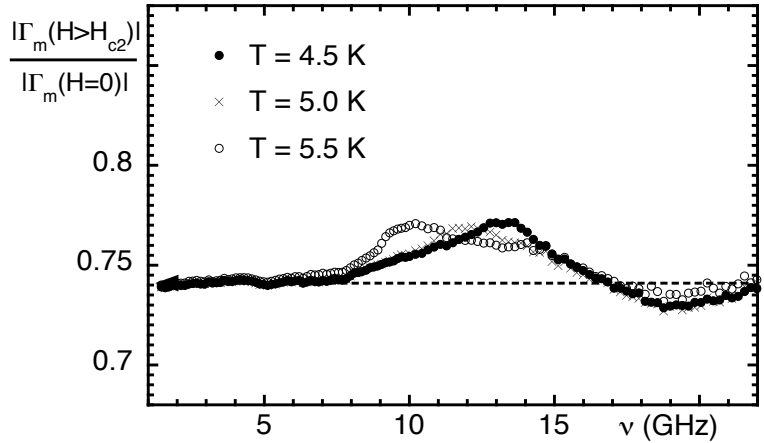}} 
\caption{Ratio  $|\Gamma_m(T,H^*,\nu)/\Gamma_m(T,0,\nu)|\approx\Gamma_{0,n}(T)$ at  various temperatures for sample Nb. $\Gamma_{0,n}$ can be estimated  precisely in the low frequency range $\nu\leq7$ GHz, where the  measured response is constant as expected. At higher frequencies  residual oscillations, within 3\%, remains after the approximate  calibration procedure.}
\label{fig:G0n}
\end{figure}

Therefore, at low frequencies the ratio  $|{\Gamma}_m(T,H^*,\nu)/{\Gamma}_m(T,0,\nu)|\simeq|\Gamma_{0,n}(T)|$,  which is a frequency independent constant. An example based on the  measurements on the Nb sample is reported in Figure \ref{fig:G0n}. It  can be seen that, apart a residual oscillation at high frequency,  that remains within 3\%, the frequency independence of $\Gamma_{0,n}$  holds.

%

Since $\Gamma_{0,n}$ is a frequency-independent quantity, it  is possible to estimate it by using low frequency data only,  $\nu\leq7$ GHz. It is then possible to determine $\alpha(T,\nu)$, the important frequency-dependent cable coefficient, which contains  $\Gamma_{0,n}$ as a proportionality constant, according to:
%
\begin{equation}
     \alpha(T,\nu)={\Gamma}_m(T,H^*,\nu)/\Gamma_{0,n}(T)
\label{extAlpha}
\end{equation}
\noindent and then used to extract $\Gamma_0(T,H,\nu)$ for each set  of measurements corresponding to a field sweep at fixed $T$. Finally,  by using \eqref{Zs} the frequency-dependent resistivity can be  obtained. In the present measurements there is the additional  advantage that the normal state resistivity $\rho_n$ at low  temperature of our samples is not temperature dependent. Thus,  $\Gamma_{0,n}$ does not depend on temperature either and once  obtained at one temperature the value can be used to derive $\alpha$  at all $T$.
One should remark that this kind of calibration  is feasible only on thin films, where $\Gamma_{0,n}$ is  frequency-independent: the same procedure cannot be applied to the  study of bulk materials (e.g., single crystals), being in this case  $\Gamma_{0,n}$ dependent on the frequency. In this case, one should  develop an alternative, sample-in-situ, calibration.

Using \eqref{Zs} and \eqref{extAlpha} one finally obtains for the  sample complex resistivity $\rho(\nu)$:
\begin{equation}
     \label{rho1finale}
     \rho_1(\nu)\simeq 
Z_{0}d\frac{1-\left|\frac{{\Gamma}_{m}(T,H,\nu)}{{\Gamma}_{m}(T,H^*,\nu)}\right||\Gamma_{0,n}|}{1+\left|\frac{{\Gamma}_{m}(T,H,\nu)}{{\Gamma}_{m}(T,H^*,\nu)}\right||\Gamma_{0,n}|}
\end{equation}
\begin{equation}
     \label{rho2finale}
     \rho_2(\nu)\simeq 
-2Z_{0}d\frac{\left|\frac{{\Gamma}_{m}(T,H,\nu)}{{\Gamma}_{m}(T,H^*,\nu)}\right||\Gamma_{0,n}|}{\left(1+\left|\frac{{\Gamma}_{m}(T,H,\nu)}{\Gamma_m(T,H^*,\nu)}\right||\Gamma_{0,n}|\right)^2}\Delta\phi_{m}
\end{equation}
\noindent where we have used the approximations  $\Re\left(\frac{{\Gamma}_{m}(T,H,\nu)}{{\Gamma}_{m}(T,H^*,\nu)}\right)\simeq  \left|\frac{\Gamma_m(T,H,\nu)}{\Gamma_m(T,H^*,\nu)}\right|$ and  $\Im\left(\frac{\Gamma_m(T,H,\nu)}{\Gamma_m(T,H^*,\nu)}\right)\simeq  \left|\frac{\Gamma_m(T,H,\nu)}{\Gamma_m(T,H^*,\nu)}\right|\Delta\phi_{m}$  [where $\Delta\phi_{m}=\angle\Gamma_m(T,H,\nu)-\angle\Gamma_m(T,H^*,\nu)$],  which hold true since $\Delta\phi_{m}$ is small. These expressions clearly show that the real part of the resistivity  can be obtained solely by the modulus of the reflection coefficient,  whereas the imaginary part strongly depends on the phase. On  practical grounds, this is a relevant remark: whenever thermal or  other instabilities affect the cable response, they impact largely  and mainly on the measurement of the phase rather than on the modulus:  in these cases, the above expressions show that although the  imaginary part of the resistivity would be largely affected, the real  part would still remain reliably determined. In the measurements here  presented we limit ourself to the extraction of the more reliable  real part of the resistivity, $\rho_{1}$.
\section{Experimental results}
\label{exp}
In this Section we present experimental results for the  frequency-dependent real part of the resistivity in the mixed state  in three superconducting samples: a pure Nb film of 20 nm nominal  thickness and two Nb/Pd$_{84}$Ni$_{16}$/Nb trilayers. The samples  were fabricated at University of Salerno/CNR-SPIN. They were grown by  ultra-high-vacuum, dc magnetron sputtering. The films and  heterostructures were grown at room temperature, and the thickness  was carefully calibrated against deposition rate, following the  procedure described in more details elsewhere \cite{IlyinaPhC10}.  Since microwave measurements  can be seriously distorted by the use  of semiconducting substrates, as extensively discussed previously  \cite{pompeoSUST05}, samples were deposited on crystalline sapphire  (Al$_2$O$_3$) instead of the more common choice of Si. We focused the  attention on samples with small $d_F$, in order to assess whether  ferromagnetic layers thinner than the coherence length in the  ferromagnet, $\xi_F$, can give rise to detectable features in the  transport properties, apart the well-known drop in $T_c$  \cite{cirilloPRB09}. Since in Nb/PdNi/Nb one estimates $\xi_F\approx$  3-4 nm \cite{cirilloPRB09}, we have considered here the following  three samples: a Nb thin films (sample Nb), with nominal thickness  $d=$20 nm, as a reference, and two trilayers with nominal Nb  thickness $d_S=$ 15 nm for upper and lower layers and PdNi thickness  $d_F=$ 1 nm and 2 nm (samples T1 and T2, respectively). The normal  state dc resistivity was in all cases $\rho_n=25\pm 5\; \mu\Omega$cm.  The main characteristic parameters are reported in the Table. The  critical temperature $T_c$ and the upper critical fields $H_{c2}$  were estimated from the disappearance of the microwave signal, and  were found in good agreement with the dc data. The magnetic field $H$  was applied perpendicular to the film plane.

As described in Section \ref{tech}, by measuring the changes with the  applied magnetic field $H$ of the frequency-dependent complex  reflection coefficient of the samples we derived the field-increase  of $\rho_1$ on the basis of the Equation \eqref{rho1finale}. Apart  from a very narrow region in the $H-T$ phase diagram close to the  transition line $H_{c2}(T)$, we do not expect a detectable  contribution of pairbreaking at our working frequency, so that  we  entirely ascribe the measured signal to the vortex resistivity. We  anticipate that the frequency dependence of the measured real  resistivity is well described by the general theoretical expression  \eqref{eq:rhovm}, with no need for other contributions, thus adding  confidence in the assumption. For this reason, when dealing with  experimental quantity we will use the symbol $\rho_{vm,1}$ for the  experimental field increase of the real resistivity. In order to  directly compare different samples, we report in the following the  normalized resistivity, $\rho_{vm,1}/\rho_n$.

Measurements have been taken approximately at the same reduced  temperature $t=T/T_c\approx 0.72$ to compare similar thermal  properties, and at the same magnetic fields $\mu_0 H=\;$0.6 and 0.8 T  in order to have the same vortex density. Sufficiently large magnetic  fields have been chosen in order to examine also the creep factor in  the three samples, which becomes negligibly small at lower fields.

\begin{figure}[h]
\centerline{\includegraphics[width=10cm]{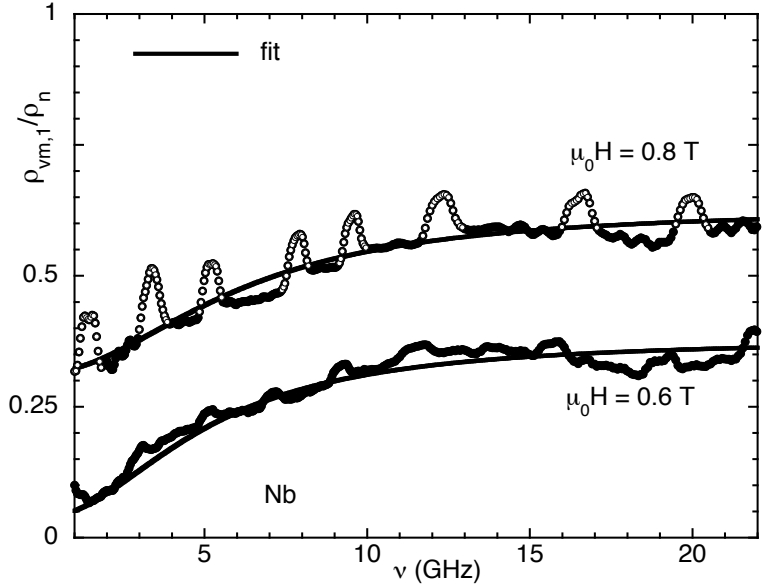}}
   \caption{Normalized vortex resistivity $\rho_{vm,1}/\rho_n$ as a  function of the frequency $\nu$ in the Nb sample, at two different  magnetic fields and $T=\;$4.5 K. Symbols: experimental data; open  symbols: data in the ``switched" state (see text). Continuous lines:  fits with the general expression \eqref{eq:rhovm}, with parameters  reported in the Table.}
   \label{figNb}
\end{figure}

In Figure \ref{figNb} we report $\rho_{vm,1}$ in the Nb sample taken  at $T=\;$4.5 K. We first notice the appearance of steps in  $\rho_{vm,1}$ at the larger field. The resistivity switches toward a  more dissipative (larger $\rho_{vm,1}$) state and then decays  abruptly, then switches again.
We remark that, to the best of  our knowledge, such instabilities have not been reported before. In  particular, a time-domain analysis reveals that the switching  phenomena here reported have a very long typical time scale, $\sim$1s  \cite{pompeoPhC10}.
We will not focus further on this anomalous  behaviour, that is presently under study. We mention that it is  invariably present in our measurements, although somehow  sample-dependent, it is generally more evident approaching the  superconducting transition\footnote{While this is a general trend, a  few exceptions exist. For example, in the measurements of Figure  \ref{figT2}, taken close to $H_{c2}$, no switches are evident.} 
and completely disappears in the normal state (i.e., for  $H>H_{c2}$ or $T<T_c$). This behaviour has been verified with both  VNAs.
 Further details and a characterization of the phenomenon have  been given elsewhere \cite{pompeoPhC10}. We focus instead on the  ``conventional" behaviour, indicated in Figure \ref{figNb} by full  symbols, that we compare to the theoretical expression  \eqref{eq:rhovm}.
Directly from the data it can be seen that the vortex characteristic  frequency lies well within our measuring range of 1-22 GHz, as  indicated by the steep rise in $\rho_{vm,1}/\rho_n$. As a  consequence, we are able to observe the free-flux-flow plateau in  $\rho_{vm,1}$ at the high edge of our measuring range. A quantitative  fit with \eqref{eq:rhovm} is reported in Figure \ref{figNb}. The fit  is in good agreement with the data, and it gives the value for the  vortex parameters reported in the Table.

We note the rather high value of the characteristic frequency,  $\nu_0=\;$5 and 6 GHz at $\mu_0H=\;$0.6 and 0.8 T, respectively. This  result is consistent with the depinning frequency in Nb thin films as  measured with single-frequency resonators \cite{JanjusevicPRB06}, and  it indicates quite a strong pinning in our Nb film. We also note that  the normalized flux-flow resistivity $\rho_{ff}/\rho_n<H/H_{c2}$ by a  significant factor. This remark is a direct demonstration that the  simple Bardeen-Stephen model \cite{BS}, where the flux flow  resistivity $\rho_{ff}=c\rho_n H/H_{c2}$ with $c\simeq$1, does not  describe well our data. While it should be mentioned that, when a  full time-dependent Ginzburg-Landau calculation is performed,  $\rho_{ff}$ can be much smaller than the BS value \cite{kopninBOOK},  it seems that this result, which we recall is independent from a  particular model for vortex motion, calls for further studies of the  flux flow resistivity in conventional superconductors. Measurements  are presently in progress.

Finally, we remark that a significant creep exists at the higher  measuring field, $\epsilon_{eff}(0.8\;$T$)=0.5$. This is an  interesting feature, since the role of creep is often neglected in  the analysis of high-frequency data in conventional superconductors  \cite{JanjusevicPRB06}, under the assumption that it is relevant only  extremely close to the transition. We notice that recent  low-temperature, multifrequency microwave measurements in  conventional superconductors revealed some contribution from flux  creep \cite{songPRB09}.
It appears that multifrequency or, like in  the present work, swept-frequency measurements are a particularly  suitable tool to directly uncover features in the vortex state  resistivity.  
We propose a framework in order to explain the pinning and creep features found in the present measurements. First, one should consider that at microwave frequencies the flux lines undergo very small displacements around their equilibrium position. In this sense, the  pinning probed by the microwave response concerns the curvature at the bottom of the potential well, and has (as a first approximation) nothing to do with the height of the pinning well itself. This is a significant difference with respect to conventional dc measurements and pinning, as measured by the critical current density $J_c$, is essentially dictated by the height of the potential wells. By contrast, creep is determined by the height of the wells both at microwaves and in dc. The present measurements can then be explained in terms of rather steep, but not very deep pinning wells: the pinning-related parameter $\nu_0$, being rather high, points to steep pinning wells, while the large creep factor indicates that the pinning wells themselves are not particularly deep. A check of this framework comes from the magnetic field dependence of the vortex parameters: increasing the field should first reduce the height of the pinning wells, and only as a further effect it should affect the shape of the bottom of the well. This behaviour should result in a quickly increasing creep factor with the field, and a weakly field dependent depinning frequency (until $H_{c2}$ is approached, where the pinning potential flattens and the characteristic frequency vanishes). In fact, it can be seen in the Table that with increasing field $\nu_0$ does not drop quickly, while $\epsilon_{eff}$ increases significantly (anticipating the results obtained in the trilayers, we observe the same trend, even if  $H_{c2}$ and pinning are smaller in sample T2). The physics at high driving frequency differs from the better known flux motion processes in dc: at low dc currents the motion of flux lines arises as flux bundles \cite{blatter}, which require quite a large activation energy. Hence, the small creep factor as seen in dc unless the current is raised. By contrast, the motion at microwave frequencies is much closer to the single-fluxon motion, as shown by the study of the elastic force matrix for the flux lattice \cite{ongPRB97}: with increasing frequency, the range of the transmitted forces between fluxons drops thus yielding the single-vortex regime. In this case, the activation energies related to single vortex creep are smaller \cite{blatter}.

Having determined the vortex parameters in Nb, we turn to the  analysis of the data in the trilayers.
\begin{figure}[h]
\centerline{\includegraphics[width=10cm]{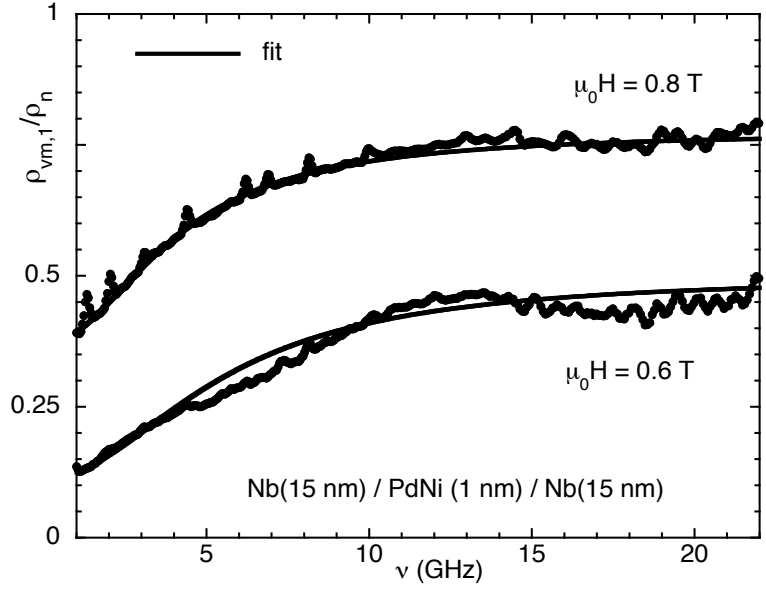}}
   \caption{Normalized vortex resistivity $\rho_{vm,1}/\rho_n$ as a  function of the frequency $\nu$ in the trilayer T1, with 1 nm PdNi  thickness, at two different magnetic fields and $T=\;$4.5 K. Symbols:  experimental data. Continuous lines: fits with the general expression  \eqref{eq:rhovm}, with parameters reported in the Table. The  frequency dependence of $\rho_{vm,1}$ is very similar to analogous  data for the reference Nb sample (Figure \ref{figNb}).}
   \label{figT1}
\end{figure}

Figure \ref{figT1} reports $\rho_{vm,1}/\rho_n$ in sample T1,  together with the fits with \eqref{eq:rhovm}. It turns out that  sample T1, where the PdNi thickness is limited to $d_F=\;$1 nm, has a  behaviour very close to the Nb film for what concerns the vortex  resistivity. The effect of the F layer seems limited to the reduction  of $T_c$ by about 1.2 K with respect to a single Nb film of thickness  30 nm \cite{cirilloPRB09}. The vortex parameters are reported in the  Table. Again, direct indications of strong pinning and significant  creep is found, and $\rho_{ff}/\rho_n<H/H_{c2}$. The same  considerations already stated for Nb are valid. It seems that  $d_F=\;$1 nm is not sufficient to affect the vortex state resistivity  in our S/F/S trilayers.
\begin{figure}[h]
\centerline{\includegraphics[width=10cm]{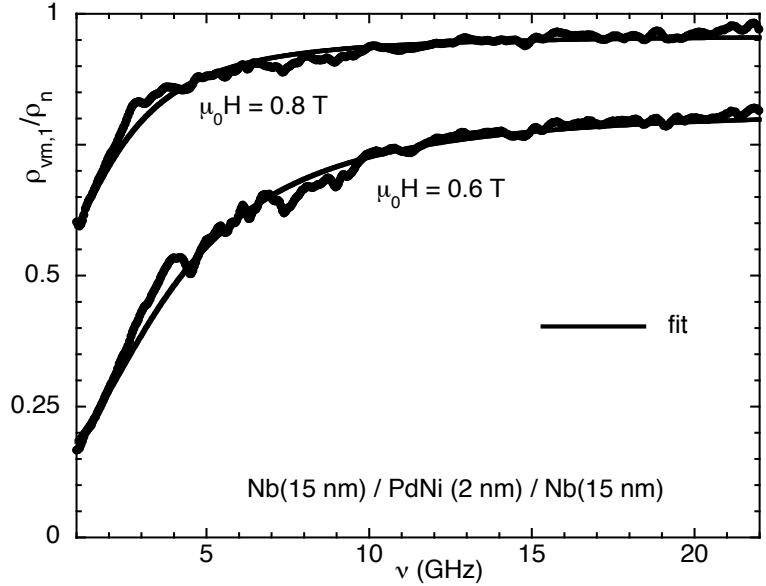}}
   \caption{Normalized vortex resistivity $\rho_{vm,1}/\rho_n$ as a  function of the frequency $\nu$ in the trilayer T2, with 2 nm PdNi  thickness, at two different magnetic fields and $T=\;$3.66 K.  Symbols: experimental data. Continuous lines: fits with the general  expression \eqref{eq:rhovm}, with parameters reported in the Table.  It can be directly seen from the data that the frequency dependence  of $\rho_{vm,1}$ is different with respect to the reference Nb sample  and to the trilayer with $d_F$= 1 nm  (Figures \ref{figNb} and  \ref{figT1}, respectively).}
   \label{figT2}
\end{figure}

The data for sample T2 are presented in Figure \ref{figT2}, again  with the fits obtained using \eqref{eq:rhovm}. Even if the thickness  of the F layer is only 2 nm, significant differences with Nb and  trilayer T1 appear. First, pinning is significantly reduced, as  evident from the much steeper raise in $\rho_{vm,1}/\rho_n$.  Performing the fits, one finds that the characteristic vortex  frequency is reduced by an appreciable amount with respect to Nb and  trilayer T1, while creep does not change much. However, the most  remarkable feature is the value of the flux flow resistivity. In this  case we find $\rho_{ff}/\rho_n>H/H_{c2}$, which is \textit{above} the  BS value. This finding cannot be easily explained, and clearly calls  for further investigations. It should be stressed that the dynamics  remains unchanged, meaning with this statement that the frequency  dependence is still the same of pure Nb and of sample T1,  satisfactorily described by \eqref{eq:rhovm}. Within the wide class  of models represented by \eqref{eq:rhovm}, the high values of  $\rho_{ff}/\rho_n$ must find an explanation in the structure of the  vortex itself, and/or in the quasiparticle relaxation processes in  the superconducting state in presence of a magnetic field, since no  pinning or thermal effects can be invoked. A study of the vortex  resistivity as a function of $d_F$ is clearly required, and it will  be the subject of ongoing investigation.
\begin{table}
\caption{\label{table}Main parameters for the samples under study.}

\begin{indented}
\lineup
\item[]\begin{tabular}{@{}*{7}{lccc}}
\br
\0  & Nb & T1 & T2 \cr
\mr
\0\0Nb thickness (nm) $^{(1)}$ & 20 & 15 & 15 \cr
\0\0PdNi thickness (nm) $^{(1)}$ & --- & 1 & 2 \cr
\0\0$T_c$ (K) $^{(2)}$ & 6.2$\pm$0.1 & 6.2$\pm$0.1 & 5.1$\pm$0.1 \cr
\0\0Measuring $T_m$ (K) $^{(3)}$ & 4.50$\pm$0.01 & 4.50$\pm$0.01 & 3.66$\pm$0.01 \cr
\0\0$\mu_0 H_{c2}(T_m)$ (T) $^{(2)}$ & 1.00$\pm$0.05 & 0.90$\pm$0.05 & 0.85$\pm$0.05 \cr
\br
\0\0vortex parameters$^{(4)}$ at $\mu_0H=\;$0.6 T &   &   &   \cr
\0\0$\rho_{ff}/\rho_n$ & 0.38$\pm$0.02 & 0.50$\pm$0.02 & 0.82$\pm$0.02 \cr
\0\0$\nu_{0}$ (GHz) & 5.0$\pm$0.4 & 5.5$\pm$0.5 & 4.0$\pm$0.4 \cr
\0\0$\epsilon_{eff}$ & 0.10$\pm$0.02 & 0.23$\pm$0.03 & 0.17$\pm$0.03 \cr
\br
\0\0vortex parameters$^{(4)}$ at $\mu_0H=\;$0.8 T &   &   &   \cr
\0\0$\rho_{ff}/\rho_n$ & 0.63$\pm$0.02 & 0.78$\pm$0.01 & 0.96$\pm$0.01 \cr
\0\0$\nu_{0}$ (GHz) & 6.0$\pm$0.4 & 4.0$\pm$0.4 & 2.3$\pm$0.3 \cr
\0\0$\epsilon_{eff}$ & 0.50$\pm$0.01 & 0.48$\pm$0.03 & 0.55$\pm$0.03 \cr
\br
\end{tabular}
\end{indented}
\centerline{\tiny $^{(1)}$ Nominal values, as determined through careful calibration against deposition rate \cite{IlyinaPhC10}.}
\vspace{-3mm}
\centerline{\tiny $^{(2)}$ Estimated from the disappearance of the microwave signal.}
\vspace{-3mm}
\centerline{\tiny $^{(3)}$ Errors report the temperature stability during each measurement.}
\vspace{-3mm}
\centerline{\tiny $^{(4)}$ Uncertainties estimated through the sensitivity of the fit quality to each fitting parameter.}
\end{table}

\section{Conclusions}
\label{conc}
In this paper we have presented the first measurements of the  frequency dependence of the vortex-state microwave resistivity in  S/F/S trilayers, with Nb as S layer and PdNi as F layer, and on a Nb  sample. We have used a wideband technique, the Corbino disk  technique, capable to obtain reliable measurements in the frequency  range 1-22 GHz. In Nb, significant creep is detected and a previously  unobserved ``switching" behaviour is found approaching the transition  line. In trilayers the experimental data show that, even when the F  thickness $d_F$ is as thin as half of the coherence length in the  ferromagnet, $\xi_F$, the superconducting state changes from the pure  superconductor in a way that the vortex state resistivity is  affected. By exploiting the wideband technique, we have shown that  changes in the free-flux-flow resistivity are the most relevant,  since one obtains $\rho_{ff}$ in excess of the BS limit $H/H_{c2}$.  In addition, one observes a reduced pinning. By contrast, the  trilayer with $d_F$ = 1 nm does not exhibit any qualitative  difference with respect to the Nb reference sample. Our observations  call for further study of the vortex state resistivity in S/F/S  heterostructures, which are presently in progress.
\ack{We thank E. A. Ilyina, C. Cirillo and C. Attanasio for sample  preparation and useful discussions. This work has been partially  supported by the Italian MIUR-PRIN 07 project ``Propriet\`a di  trasporto elettrico dc e ac di strutture ibride stratificate  superconduttore/ferromagnete realizzate con materiali tradizionali".}

\end{document}